\documentclass{article}

\usepackage[margin=1in]{geometry} 

\usepackage{graphicx} 
\usepackage{parskip}
\usepackage{amsmath}
\usepackage{algpseudocode}
\usepackage{booktabs} 
\usepackage{url}
\usepackage{hyperref}
\usepackage{authblk}

\usepackage[utf8]{inputenc}
\usepackage{tabularx}
\usepackage{multirow}

\usepackage{xcolor,colortbl} 

\definecolor{Gray}{gray}{0.85}
\definecolor{LightCyan}{rgb}{0.88,1,1}
\definecolor{MinusBlue}{rgb}{1,1,0.88}
\definecolor{MinusGreen}{rgb}{1,0.88,1}
\definecolor{LightRed}{rgb}{1,0.5,0.5}

\newcolumntype{b}{>{\columncolor{LightCyan}}c}
\newcolumntype{a}{>{\columncolor{MinusBlue}}c}
\newcolumntype{d}{>{\columncolor{MinusGreen}}c}
\newcolumntype{e}{>{\columncolor{LightRed}}c}

\title{Bayesian Optimal Phase II design with optimised stopping boundaries and response-adaptive randomisation. }

\author[1,*]{Connor Fitchett}
\author[2]{Ayon Mukherjee}
\author[1]{Sof\'{i}a S.\ Villar}
\author[1]{David S.\ Robertson}

\affil[1]{MRC Biostatistics Unit, University of Cambridge}
\affil[2]{Population Health Sciences Institute, Newcastle University}
\affil[*]{\small{Corresponding email: connor.fitchett@mrc-bsu.cam.ac.uk}}

\date{}

\begin{document}

\maketitle


\begin{abstract}
    The Bayesian Optimal Phase II (BOP2) framework is a flexible trial design that can naturally facilitate complex adaptations due to its Bayesian setting. BOP2 uses equal randomisation and equally placed interim analyses in its design, but it is unclear whether these give the best operating characteristics. By incorporating Bayesian Response-Adaptive Randomisation (BRAR) and optimal interim analysis placement, we show that allocation to the best treatment and expected sample size can be improved with minimal impact on power. We discuss recommendations on implementing
    these adaptations, using simulation-based evidence, to give practical advice to practitioners. Reproducible code for the simulations is freely provided.    \\

\end{abstract}

\section{Introduction}

Phase II clinical trials are typically used to identify whether a drug has promising efficacy: enough to proceed to a phase III design and eventually to market. They can evaluate different number of treatment arms: single-arm, examining only the experimental treatment; two-arm, which compares the treatment against an accepted standard of care (a control); or multi-arm, where several drugs may be evaluated at once in order to save time and money. Single-arm trials were historically common, but early phase two-arm and multi-arm studies are becoming more popular. This could be in part due to type I error rate inflation issues in single-arm designs \cite{tang2010}, but also in part due to the US Food and Drug Administration (FDA) new Project Optimus guidance, which encourages more work on early phase dose finding to identify the best dose for the patient \cite{optimus}. Project Optimus aims to move past just accepting the maximum tolerated dose (MTD) and instead promotes evaluating both safety and efficacy in order to find the dose that is best for the patient. 

This shift in focus towards flexible endpoints that can model any treatment has paved the way for novel trail designs such as BOP2. BOP2 (Bayesian Optimal Phase II) is a trial design originally proposed by Zhou et al.
(2017) \cite{zhou2017bop2}, designed for versatility to innately handle any endpoint that can be modelled in a Bayesian setting. At interim analyses, it uses Bayesian posteriors to determine the probability that one treatment is better than another, and uses this to decide whether the trial should be ended early for futility or superiority. It is optimal in the sense that it selects its stopping rules to protect the type I error rate whilst maximising the power for pre-selected null and alternate hypothesises. 

Since its implementation, it has been used in over 50 trials at MD Anderson \cite{zhao2023bop2} and has been the focus of several papers extending its reach. Zhao et al. (2023) \cite{zhao2023bop2} introduces BOP2-DC (BOP2 Dual-Criterion) which extends decision making from simple rejecting/accepting a treatment to rejecting/ accepting/ consider, signifying when it is best to gather more information rather than end the trial with an uncertain decision. Ding (2023) \cite{Ding2023} considers Randomised-BOP2, where one of the main advantages is that it does not limit the number of interim analyses, thus allowing as many checks as would be suitable for the trial in progress. Zhou et al. (2020) \cite{Zhou2020BOP2} adapts the Bayesian model of BOP2 so that it caters to time-to-event data such as PFS (Progression Free Survival), giving practitioners more choice of what endpoints to evaluate. Of most note to this paper however are  Zhao et al. (2022) and Mulier et al. (2024) \cite{Zhao2022} \cite{Mulier2024}, which extended the one-arm BOP2 to the two-arm and multi-arm settings, respectively. Trials beyond one-arm allow us to consider a wider range of adaptations, such as modifying how patients are allocated. Thus this paper will be taking these designs and considering what benefits can be gained by modifying the randomisation and early stopping regimes. 

BOP2 already incorporates early stopping rules but assumes that the interim analyses are evenly spaced. We will going further by analysing what benefits could be gained if we optimised the number and timing of interim analyses (IAs). Early stopping by itself can already reduce the average number of patients that need to be enrolled in a trial, but several papers show that optimising IA timing can lead to improved patient benefit metrics, across a range of binary and continuous endpoints. Simon (1989) \cite{simon1989} proposed an optimal two stage design which minimises the expected sample size (ESS) and is now commonly referenced in the trial design literature. The optimisation works via enumeration, calculating every single IA timing and then calculating what is the best ESS. Building on this work, Feng and Zee (2024) \cite{feng2024} attempt to move on from enumeration, as it can be computationally taxing, and introduce a utility function that effectively combines power and ESS. Whilst utility functions do allow a compromise between ESS and power, they are inevitably held back by the fact that the utility weight needs to be pre-specified. Another theoretical approach is the curtailment approach of Brannath et al. (2006) \cite{brannath2006}. This allows the practitioner to be flexible in the number of IAs they wish to consider, as there is the option to remove a future IA at a current IA if you would be unlikely to make a decision there. Finally, both Jung et al. (2001) and Fleming (1982) \cite{jung2001} \cite{fleming1982} optimise the IA via enumeration using different criterion to Simon (1989) \cite{simon1989} rather than simply minimising the ESS. Qin et al. (2020) \cite{qin2020} compares the three methods, evaluating when each method performs best. This paper will generally focus on enumeration techniques to optimise ESS as well as extending this enumeration work to multi-arm trials, with most discussion on this being found in Section \ref{sec:IA}.

Alongside having direct effects on patient benefit metrics, the timing and frequency can also effect the potential value of any response-adaptive randomisation (RAR) scheme that is used. BOP2 in its original papers has fixed, equal allocation to all of its treatment arms; an aim of this paper is to illustrate that patient benefit metrics can be improved when RAR is implemented thoughtfully. RAR moves the randomisation away from fixed randomisation (typically equal randomisation amongst all the arms, although other schemes exist) and towards an adaptive scheme that learns from the data: as more patient data are gathered, most types of RAR will skew the allocation probabilities so that more patients are allocated towards the best performing arm. Robertson et al. (2023) \cite{robertson2023} summarises a wide range of RAR which has been analysed in the current literature, including the RAR this paper focuses on: Bayesian RAR (BRAR). Initially defined by Thompson (1933) \cite{thompson1933}, BRAR uses Bayesian posterior probabilities to guide the allocation of patients, therefore naturally working well with Bayesian designs. BRAR has now been successfully used in modern trials such as BATTLE \cite{kim2011battle} and I-SPY 2 \cite{barker2009}. 

Bayesian Response Adaptive Randomisation fits into the regulatory focus by promoting efficient, ethical, and data-driven trial designs that maintain statistical rigour while enabling regulators to evaluate emerging evidence in a transparent and pre-specified framework. This concept is also well aligned with the FDA’s Project FrontRunner \cite{frontrunner} , which seeks to accelerate the development of oncology therapies in earlier treatment settings by encouraging more efficient and flexible trial designs. Under the FrontRunner initiative, sponsors are incentivized to generate robust, randomised evidence even in first- or second-line metastatic disease. BRAR complements this goal by dynamically skewing patient allocation toward more promising treatment arms based on accruing efficacy (or safety) data, thereby increasing the ethical appeal of trials, improving statistical efficiency, and potentially shortening decision timelines in line with FrontRunner’s mission.

This paper therefore aims to evaluate the effect that changing the randomisation scheme as well as the timing and frequency of IAs can have on the trial operating characteristics. We will be simulating a wide range of scenarios using different IA and RAR methods before evaluating them and making recommendations based on our results. Additionally, we will be making the contribution of considering RAR and IA in BOP2 jointly: how does the final proportion of patients allocated to RAR change as we move the IA, and can we exploit this to improve the BOP2 design. Finally, this paper also provides the reproducible code used for all of its simulations.

This paper now proceeds as follows. Firstly, Section~\ref{sec:BOP2} presents an overview of the BOP2 trial design and introduces the relevant notation. Section~\ref{sec:IA} then discusses IA timing and frequency, and the advantages that optimising IAs can bring to BOP2. Next, Section~\ref{sec:RAR} introduces BRAR and applies this to BOP2. Alongside defining the RAR schemes we will be using, this section also presents simulation results that illustrate how BOP2 changes as we alter the randomisation scheme. Section~\ref{sec:RAR} analyses the effects of jointly implementing RAR and optimal IA placement. Finally, Section 5 presents a discussion and practical recommendations to readers, as well as potential avenues for future work.

\section{BOP2 framework}
\label{sec:BOP2}

The original BOP2 design was single-arm and only accounted for early stopping due to futility \cite{zhou2017bop2}. In their binary survival endpoint example, they modelled positive patient responses (1 for success, 0 for failure) with a $\pi \sim \operatorname{Beta}(1,1)$ prior (i.e.\ a uniform prior on [0,1]) paired with a binomial likelihood, $\operatorname{Binomial}(n_{1},\theta_1)$, to model the probability of survival for the treatment, $\theta_{1}$, based on the number of patients allocated to it, $n_{1}$. Before the trial, they would specify points to carry out IA, with their stopping boundary being given by the posterior probability that the experimental treatment is worse than the current standard of care. If the probability of it being worse was sufficiently high, the trial was ended for futility:

\[ P( {\theta}_{1} \leq {\Theta} | D ) > 1 - \lambda \left(\frac{n}{N}\right)^{\gamma}. \]

Here, $\Theta$ denotes a pre-specified clinically meaningful efficacy, $D$ denotes the data collected so far, $\frac{n}{N}$ is the information fraction represented by the current sample size, $n$, over the maximum sample size $N$, and finally $\lambda$ and $\gamma$ are optimised parameters. This means that a grid search is performed over the possible values of $\lambda$ and $\gamma$ (both restricted to $[0,1]$) and the parameters which maximise the power whilst controlling the type I error rate of the trial are chosen, which then define the stopping boundaries. Roughly, as $\gamma$ increases from $0$ to $1$ the decision curve transitions from constant to linear: the more linear the more difficult it is to end for futility early on. On the other hand, $1 - \lambda$ is the final value the posterior needs to beat to reject the null hypothesis. Importantly, the Bayesian framework above is not restricted to just binary endpoints nor a specific type I error rate. The stopping rule is valid for any endpoint as long as you can pose a model where its  posterior probability is tractable. Additionally, the optimisation can be carried out to protect any target type I error rate (erroneously claim that the single arm is effective). For simplicity, this paper will follow the work  of Zhou et al. (2017) \cite{zhou2017bop2} by considering binary endpoints with a target $10\%$ type I error rate.

However, due to the limitations of single-arm trials \cite{tang2010} this paper will mainly be investigating modifications to the two-arm and multi-arm design of Zhao et al. (2023) and Mulier et al. (2024) \cite{zhao2023bop2} \cite{Mulier2024} respectively. The design of Zhao et al. (2023) \cite{zhao2023bop2} is a two-arm controlled trial, meaning it has an experimental treatment (whose success rate is $\theta_{1}$) that it will be comparing to a standard of care, a control (whose success rate is $\theta_{0}$). As before, the Bayesian prior ($\pi$) is uniform and the likelihood is a binomial. The futility stopping rule for this trial is a controlled extension, meaning we are now comparing the experimental treatment against the standard of care as a control with the boundary of Zhou et al. (2017) \cite{zhou2017bop2}:

\[  P( \theta_{1} \leq \theta_{0} | D ) > 1 - \lambda \left(\frac{n}{N}\right)^{\gamma}. \]

Additionally, this design also accommodates stopping early for efficacy, when there is overwhelming evidence to suggest that the treatment is superior to the control. This is given by

\[ P( {\theta}_{1} \geq {\theta}_{0} | D ) > 2\left[ 1 - \Phi \left( Z_{ \frac{1+\lambda}{2} } / \frac{n}{N} \right)\right] .   \] 

Here $\Phi$ is the standard normal c.d.f, and $ Z_x$ is the $x$th quantile of the standard normal, such that $\Phi ( Z_{x}) = x$.  It should be noted that the futility and efficacy stopping boundaries are different, which means there is an asymmetry to the decision making. In particular, it is easier to stop for futility when the control is outperforming the treatment than it is to stop for efficacy when the treatment is outperforming the control. The consequences of this will be illustrated in Section~\ref{sec:RAR}, but for now it aligns with the fact that the control is an accepted treatment whilst the experimental treatment is experimental: it should take more evidence to show that the treatment is superior than the control is superior. Figure~\ref{fig:Threshold_plot} in the appendix illustrates the different stopping boundaries.

For the multi-arm setting, we first need to clearly define the operating characteristics before we can interpret the decision boundaries. First, the null hypothesis will be that no arm is clinically effective (or clinically more effective than the control), $H_{0}: \theta_{i} = \theta_{0}$ , and there will be a separate alternate hypothesis for each arm: that this treatment arm is clinically effective (or clinically more effective than the control), $H_{i}: \theta_{i} > \theta_{0}$. This leads onto the definition of family wide error rate (FWER) and least power. FWER, analogous to type I error, is the probability that any arm rejects the null hypothesis when no arm is effective. Least power, analogous to power, is the probability of rejecting the null hypothesis for an arm with an effective treatment when every other arm is ineffective. Then similarly to the designs above, the decision boundaries are optimised to protect FWER (at $10\%$) whilst optimising least power.

In the multi-arm design, Mulier et al. (2024) \cite{Mulier2024} only considers futility stopping and uses the same futility boundary as above: $P( \theta_{i} \leq \theta_{0} | D ) > 1 - \lambda \left(\frac{n}{N}\right)^{\gamma}$. In addition to this, they do evaluate other efficacy boundaries that attempt to accommodate for multiple arms making erroneous stopping more likely, such as 

\[ P( \theta_{i} \leq \theta_{0} | D ) > 1 - \left(\frac{K + 1 - a - \lambda}{K + 1 - a} \right)\left(\frac{n}{N}\right)^{\gamma}. \]

Where $K$ denotes the maximum number of arms, $a$ the number of active arms that are still open to recruitment, and $i$ refers to the $i$th treatment. $K + 1 - a$ therefore shrinks as more arms drop, making it easier to end for futility. For the purpose of this paper however, we will just be considering the decision boundaries described in Zhao et al. (2022) \cite{Zhao2022}. Showcasing how BOP2 can handle complex endpoints, Mulier et al. (2024) \cite{Mulier2024} also considers a joint endpoints of toxicity and efficacy. With these endpoints being treated independently of one another it does not change the methodology: it just means we have the separate thresholds for efficacy and futility. In our work these thresholds are the same, but they could be tailored to reflect different weightings towards toxicity and efficacy.  

In the following sections we will be focusing on making modifications to the designs described above and using simulations to evaluate these changes. For most scenarios, optimisation of the $\lambda$ and $\gamma$ will have been carried out for that specific scenario beforehand. In the scenarios where it has not been optimised for each case, a type I error rate (or FWER) plot will be provided to showcase the consequences in not optimising. Optimising for every scenario is beyond the scope of this paper due to time constraints and the computational complexity of the optimisation process. 

We have investigated methods to reduce the complexity of the simulations to make the optimisation more feasible. Techniques such as parallelisation and the use of Rcpp \cite{eddelbuettel2011rcpp} save time, but the biggest efficiency is gained by using the the equations derived by Miller (2015) \cite{miller2015}, which provides exact equations for beta comparisons: working out the probability of one beta variable being the largest when comparing up to $4$ different independent betas. These exact equations both outspeed Monte Carlo simulations and reduce variability by removing simulation error \cite{kaddaj2024}. The implementation of exact posterior probability calculation using this formula does not seem to be widespread, despite the benefits it can offer \cite{kaddaj2024}. This paper uses these exact calculations to calculate posterior probabilities in two situations: 1) for decisions at IAs whether to continue the trial or not, and 2) for the calculations of allocation probabilities using BRAR. Further, we use permuted block randomisation for both equal and response-adaptive randomisation. At the cost of some predictability, this greatly reduced variation in the simulations. 

\section{Interim Analysis}
\label{sec:IA}

Simon (1989)'s \cite{simon1989} two stage work on enumerating over all the possible IA placements to optimise A) ESS and B) maximum sample size whilst controlling type I error rate and power is the main piece of motivating work for this section. We will be enumerating over timing and number of IA to jointly optimise ESS and power in the alternative hypothesis, starting by looking at how optimising IA seems to be invariant of trial properties such as probability of survival or maximum sample size in the two-arm setting. Building on Simon (1989)'s \cite{simon1989} work, we will also provide sensitivity analysis, showing how the operating characteristics can change as we slightly adjust the IA. We then repeat this analysis in the multi-arm setting, before concluding the chapter by optimising the frequency of IA. 

Throughout the rest of this paper, we use simulation studies to evaluate operating characteristics of trials. $\lambda$ and $\gamma$ have been optimised for a type I error rate of $10\%$ for most simulations. Where it would be computationally unfeasible to do so, $\lambda$ and $\gamma$ will be stated and type I error rate plots have been provided in the appendix to show that the type I rate error is still controlled. Simulations are carried out with $10,000$ iterations unless otherwise stated. We assume a $\pi \sim \operatorname{Beta}(1,1)$ (uniform) prior for all of our parameters. All figures can be reproduced via the code available in the appendix.

\subsection{IA Placement - Illustrative Example}

\begin{figure}[ht!]
    \caption{Plots showcasing how the ESS (top) and power (bottom)can change as we move our interim analysis (IA) placement. The confidence interval (CI) for the top plot is a bootstrap $95\%$ CI, whereas the CI for the bottom is an approximate normal $95\%$ CI. More details about the derivation of these CI can be found in the appendix. }
    \centering
    \includegraphics[width=1\linewidth, height = 1\linewidth]{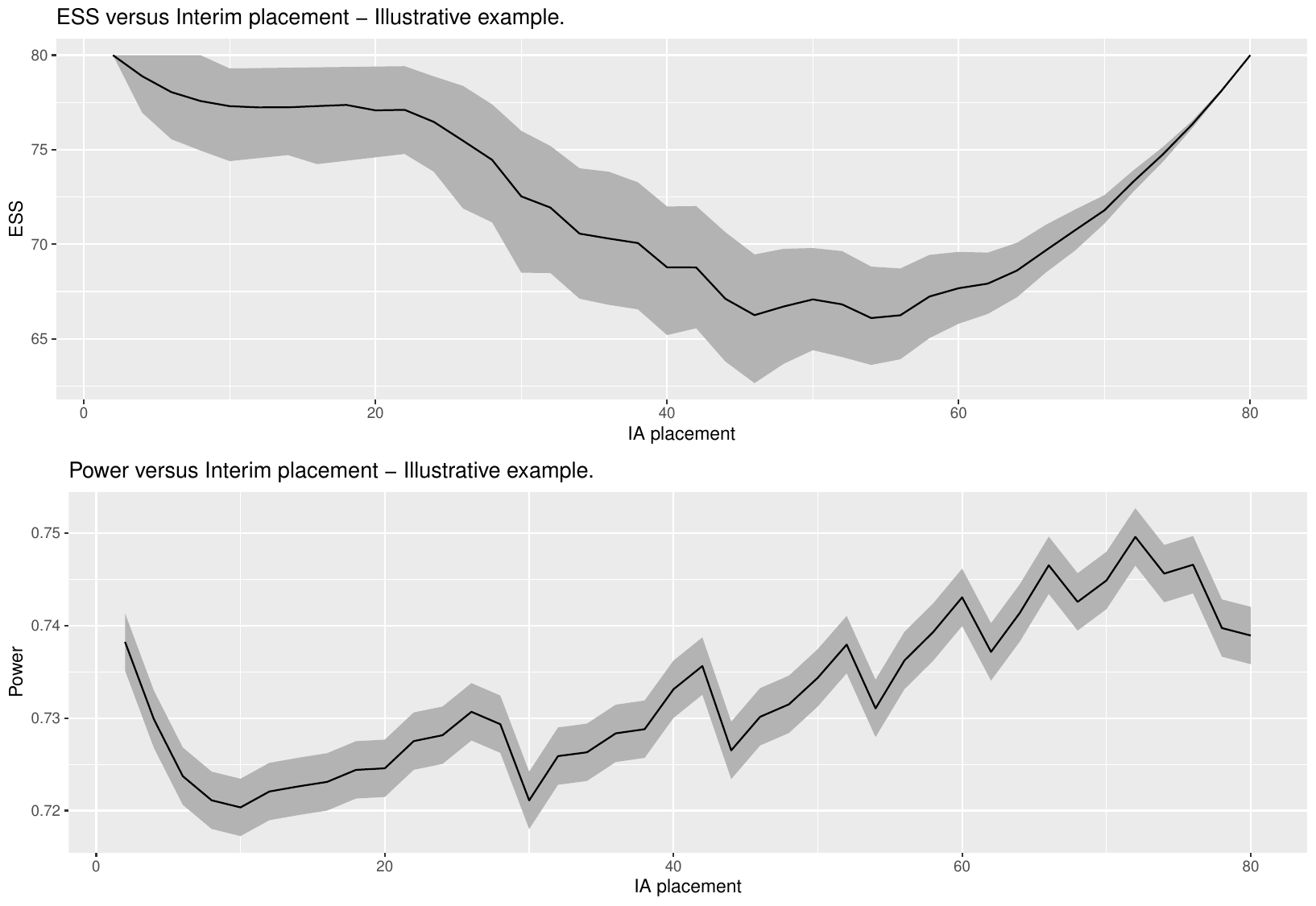}
    
    \label{fig:ESS_V_IA_plot_illustrative}
\end{figure}

First we wish to demonstrate the effect that optimising the IA can have in a two-arm trial. In Figure~\ref{fig:ESS_V_IA_plot_illustrative}, we show the effect that modifying the placement of one IA can have on the ESS and power. We assume that there is no delay between randomisation and patient response and that all patients that enter the trial are eligible for all treatments and will give full responses, yet this work could be extended to consider missing-ness or delay to endpoints. With $100,000$ simulations we consider a control success rate of $\theta_{0} = 0.2$ versus an experimental treatment success rate $\theta_{1} = 0.4$. The maximum sample size is $80$, and $\lambda = 0.91 $ and $\gamma = 0.94$ have been optimised for an IA at $60$ and a final analysis at $80$. Figure~\ref{fig:appen_2} in the appendix shows that type I error is still controlled despite not optimising for every IA placement. 

From Figure~\ref{fig:ESS_V_IA_plot_illustrative}, we can see that the optimal IA placement, in terms of minimising the ESS, is between $40$ and $60$ patients, i.e.\ between $50\%$ and $75\%$ of total patients recruited. When the ESS plateaus in a region, the power can be used to determine the which timing gives the best operating characteristics. Notice how the power curve is not smooth. This is due to the discrete nature of the data. For example, if the experimental treatment needs $5$ more successes than the control to reject the null hypothesis at a total of $30$ patients, it might also need $5$ at $31$, $32$, all the way up until $44$ where it starts needing $6$ more successes. It is more likely to get a difference of $5$ more successes in $42$ patients than in $30$, but less likely to get $6$ successes in $44$ than $5$ in $42$.  Evaluating our two plots jointly (prioritising ESS), the optimal placement for an IA would be at either $52$ or $60$ patients, as both have comparable ESS whilst being at a local maximum of the power graph.

We can also see how the effect of an IA later on in the study has diminishing effects, both not significantly reducing the ESS and reducing the power. If the IA is close to the final analysis, it becomes unlikely to make a different decision than the final analysis, hence not changing the power from a trial without an IA, whereas an earlier IA can take advantage of the power peaks we observe.

\subsection{IA Placement - Varying Trial Configuration.}

\begin{figure}[ht!]
    \caption{A pair of plots which vary the control arm success probability (whilst keeping $\theta_1 = \theta_0 + 0.2$) to see what consequence it has on the IA placement (x-axis) effecting ESS and power (y-axis).  }
    \centering
    \includegraphics[width=1\linewidth, height = 1\linewidth]{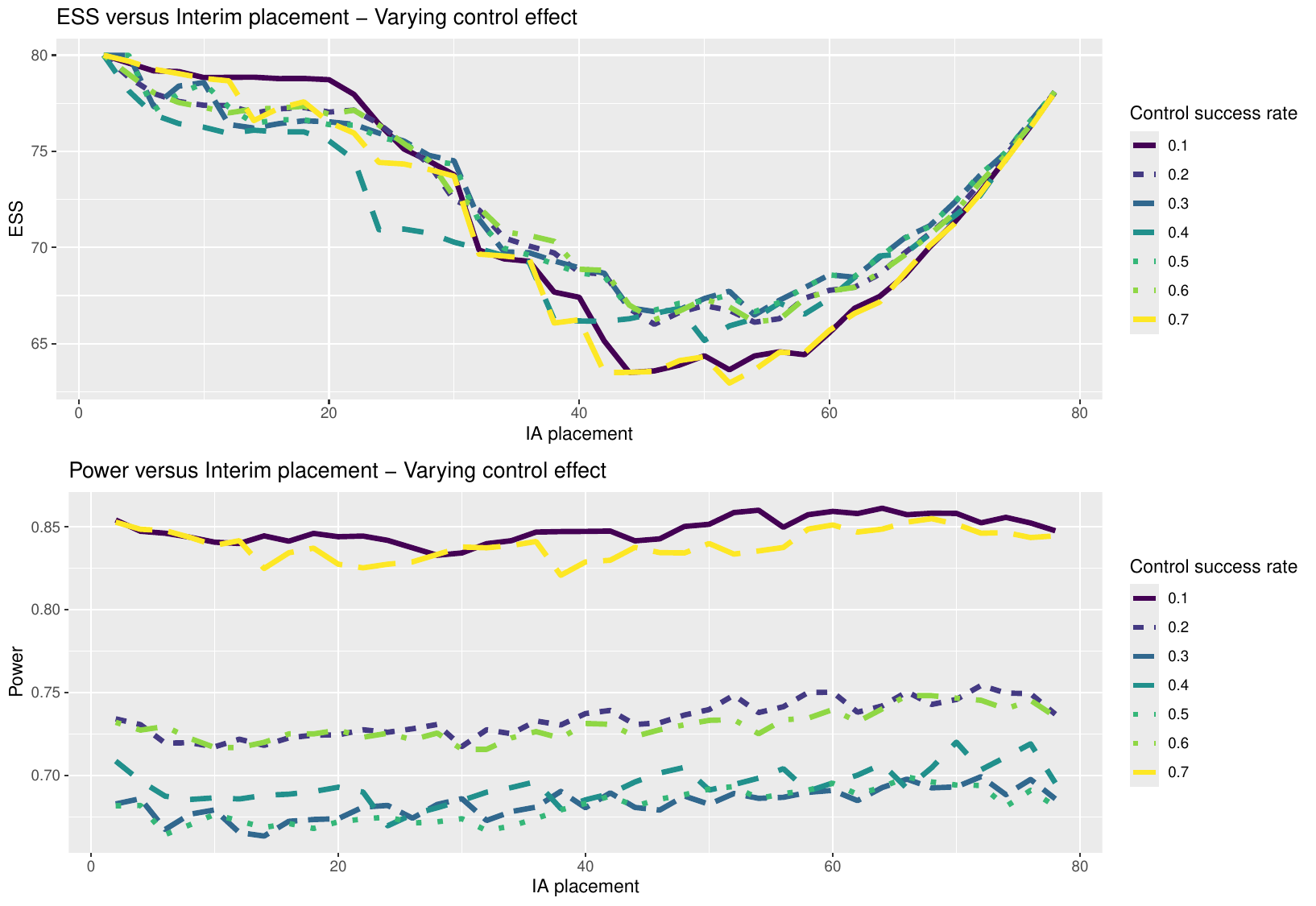}
    
    \label{fig:ESS_V_IA_plot_vary_null}
\end{figure}

\begin{figure}[ht!]
    \caption{A plot which shows the effect of IA placement on ESS across a range of maximum sample sizes. Note that the x and y axis are proportions of the maximum sample size.  }
    \centering
    \includegraphics[width=1\linewidth, height = 1\linewidth]{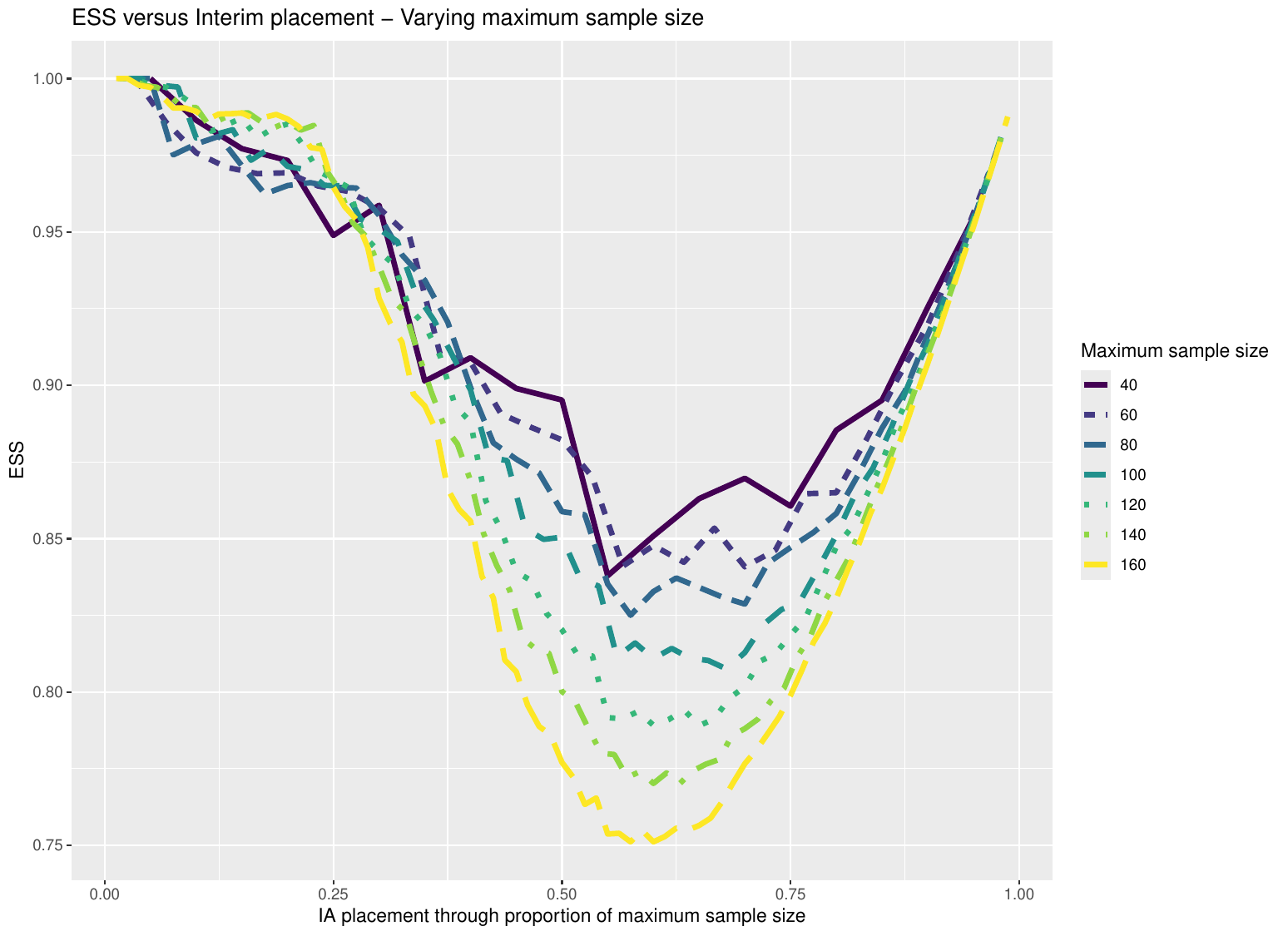}
    
    \label{fig:ESS_V_IA_plot_vary_ss}
\end{figure}

Whilst Figure~\ref{fig:ESS_V_IA_plot_illustrative} shows benefits that can be acquired from optimising, there is still uncertainty over how this generalises to other maximum sample sizes or control success rates. Figures ~\ref{fig:ESS_V_IA_plot_vary_null} and ~\ref{fig:ESS_V_IA_plot_vary_ss} showcase the effect of varying the IA timing over a range of sample sizes and control success rates, and we see similar results to Figure~\ref{fig:ESS_V_IA_plot_illustrative}. Notably, there is a plateau of ESS around $50\%$ and $75\%$ of the way through the maximum sample size.

Figure~\ref{fig:ESS_V_IA_plot_vary_null} shows plots of ESS and power vs IA placement over a range of control success rates (from $\theta_{0} = 0.1$ to $\theta_{0} = 0.7$). These were simulated with $10,000$ simulations each, and were optimised separately for each value of $\theta_0$ (exact values of $\lambda$ and $\gamma$ can be found in the code). Over this range, the relative difference between the control success rate and the experimental treatment success rate was maintained at $0.2$. 

In the power plot of Figure~\ref{fig:ESS_V_IA_plot_vary_null}, we see that the power varies as the success probabilities change. In particular, it is highest at the extreme values (when $\theta_{0} = 0.1$, $\theta_{1} = 0.3$ and $\theta_{0} = 0.7$, $\theta_{1} = 0.9$) as the variance of a Bernoulli random variable ($\theta ( 1 - \theta )$) is minimised at $\theta = 0$ and $\theta = 1$, so $\theta_{0} = 0.1$ and $\theta_{1} = 0.9$ minimise the variance, which increases the power without inflating the type I error rate. Aside from the power changes due to changing the values of $\theta_0$, we can see that the power plot is similar to Figure~\ref{fig:ESS_V_IA_plot_illustrative}, `spiky' and increasing. 

On the other hand, Figure~\ref{fig:ESS_V_IA_plot_vary_null} is qualitatively very similar to Figure~\ref{fig:ESS_V_IA_plot_illustrative} in terms of ESS. Every curve reaches an approximate minimum at around $40$, plateaus around $60$, and then rises until $80$ again. This smoothness and consistency add credence to there being theoretical justification for an optimal IA placement, corroborating the work of Feng and Zee (2024) and Rosenberger et al. (2004) \cite{feng2024} \cite{Rosenberger2004} on using utility function and constrained optimisation respectively to theoretically optimise IA. Further, this figure also corroborates the theoretical work of He et al. (2025) \cite{He2025}, which proves that the optimal ESS with a set number of IA for a general two-arm group sequential design with normally distributed endpoints is independent of the control and treatment effects: it only depends on the scheduling of the IA.

Meanwhile, in Figure~\ref{fig:ESS_V_IA_plot_vary_ss} we see that the rough shape of the ESS curve is the same but the magnitude of the trough is dependent on the maximum sample size (which varies from $40$ to $160$, once again with $\theta_0 = 0.2$ and $\theta_1 = 0.4$). This tells us that we can make greater proportional gains the larger the maximum size is. Furthermore, the minimal value again occurs at the same proportion of total patients, even with a varying sample size. Figures ~\ref{fig:ESS_V_IA_plot_vary_null} and ~\ref{fig:ESS_V_IA_plot_vary_ss} therefore demonstrate robustness in the placement of IA: that modifying the trial parameters does not greatly effect A) the optimal location of the IA or B) the effects the optimal IA brings. It is worth noting that $\lambda = 0.91$ and $\gamma = 0.98$ were only optimised for a sample size of $80$ here, but Figure~\ref{fig:appen_3} in the appendix suggests that the type I error rate is only subject to minor inflation and the results are still valid. 

\subsection{IA Placement - multi-arm.}

\begin{figure}[ht!]
    \caption{An uncontrolled multi-arm plot which shows how the ESS per arm can change as we alter the IA placement. Arm 3 is effective, whereas arms 1 and 2 are ineffective. }
    \centering
    \includegraphics[width=1\linewidth, height = 1\linewidth]{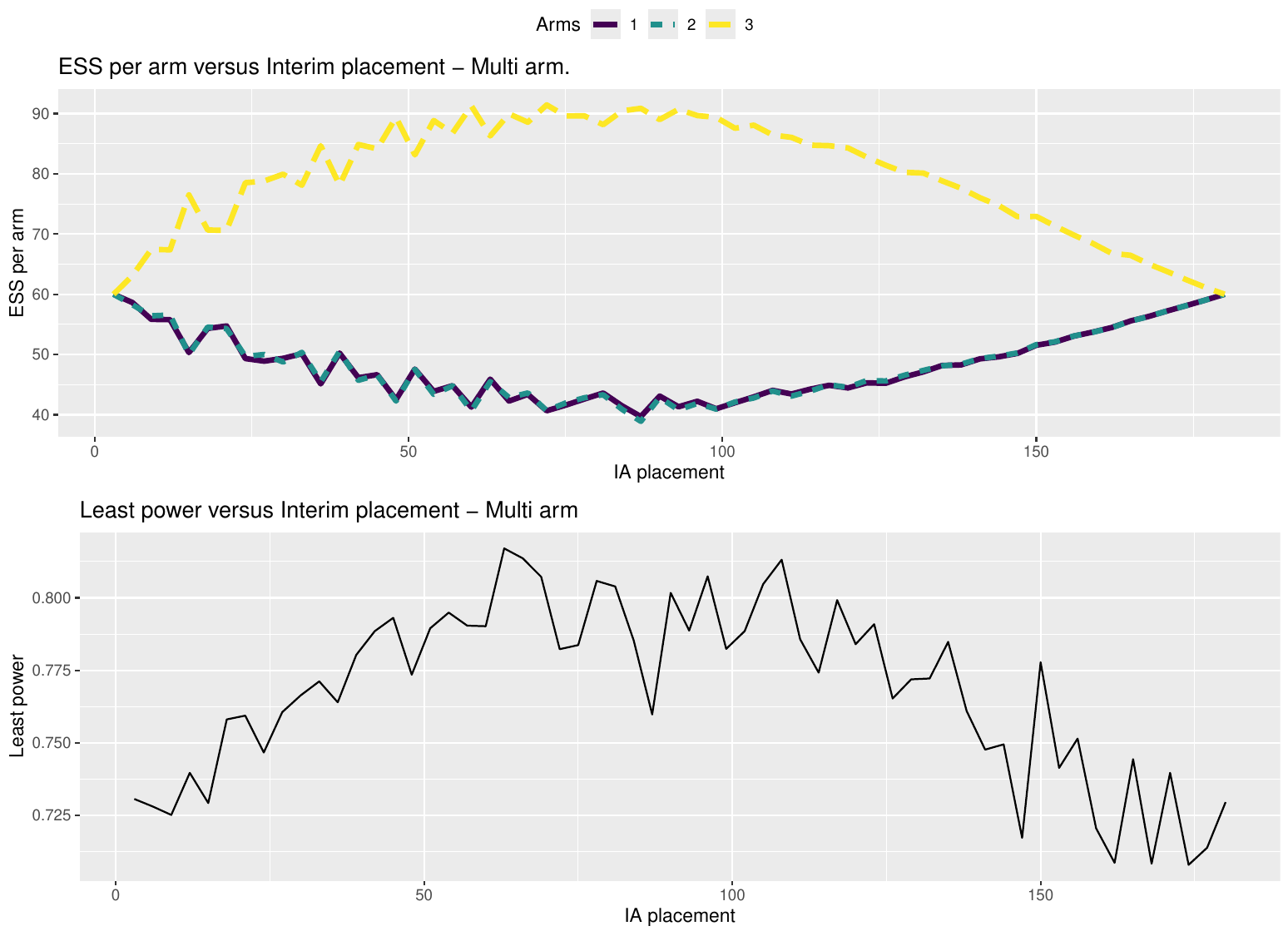}
    
    \label{fig:ESS_V_IA_plot_multi}
\end{figure}

Conducting analysis based on the ESS in multi-arm studies is more nuanced because there are two values we could be interested in. One is the total expected sample size of the trial, which we will continue to refer to as the ESS, but there is also the expected sample size in a specific arm, which we refer to as the ESS per arm. Following Mulier et al. (2024) \cite{Mulier2024}, we also have the flexibility to consider a controlled or uncontrolled trial, meaning we can choose whether to include comparisons to a standard of care control arm. The reason for considering an uncontrolled setting is for when comparing against a control may be impractical or unethical. For example, there is no accepted standard of care for a new highly lethal disease so the comparing against a placebo could be seen as an unnecessary risk of patient safety.  We also extend the design of Mulier et al. (2024) \cite{Mulier2024} in this section, by allowing patients that would have been enrolled to an arm that is dropped early to be randomised to a different, still eligible arm. This means that in this $180$ patient three arm trial, it is possible for one or two-arms to finish with more than $60$ patients each. Also like Mulier et al. (2024) \cite{Mulier2024}, we are considering joint efficacy and safety endpoints. The experimental treatment has a success rate of $\theta_{1} = 0.6$ and a toxicity rate of $\phi_{1} =  0.2$, where the two ineffective arms have a success rate of $\theta_{2,3} = 0.45$ and toxicity rate of $\phi_{2,3} = 0.3$. In order for a treatment to reject the null hypothesis, it must exceed futility and toxicity boundaries (Figure \ref{fig:Threshold_plot} in the Appendix). Additionally, recall that in a multi-arm context we are interested in FWER and least power. So in Figure~\ref{fig:ESS_V_IA_plot_multi}, $\lambda = 0.83$ and $\gamma = 0.71$ have been optimised to control FWER at $0.1$ whilst maximising least power in an uncontrolled setting (with no IA, see Figure~\ref{fig:appen_4} plot in appendix to see that the FWER is controlled). 

In Figure~\ref{fig:ESS_V_IA_plot_multi}, we can see that least power is actually improved until a point as we move our IA earlier. This is due to the reallocation adjustment, allowing running arms to exceed their initial max sample size per arm by recruiting patients from dropped arms. This allows the arms of interest to gather more patients, more data, and make more informed decisions at the end of the trial. We can also see that a well placed IA can considerably increase the ESS on the best arm, going from $60$ patients to $90$ patients in the effective arm, from recruiting $\frac{1}{3}$ to $\frac{1}{2}$ of the overall patients. This plot shows that the effect of well placed IA can have even more significant effects in a multi-arm setting. If you consider this result with the improvements to multi-arm trial characteristics in Section~\ref{sec:RAR}, the average care of a patient can be considerably improved.

\subsection{IA Frequency}

We can also consider optimising the number of IA, following in the work of McPherson (1982) \cite{mcpherson1982}. This becomes a more difficult optimisation problem however, as each IA we introduce will reduce the ESS, but by a smaller amount each time. Therefore, it is a more complicated problem then simply choosing the best ESS with an acceptable power, as this criterion may not be strict enough to choose a best number of IA. Instead, other factors such as costs and practicality of carrying out IA have to be considered and evaluated against ESS improvements to make a decision. For example, it may be impractical to organise a DMC meeting and unblind the data more than a few times for a short trial. The operational challenges that come with performing IAs can be significant but outside the scope of this paper. Instead, this paper offers quantitative guidance on the benefit that can be gained from having and optimising for more IAs, presented in Table 1.

\begin{table}[ht!]
    \caption{Table showing how the ESS can change when we place our IA using different methods. The waiting period represents a period at the start of the trial where no IA is placed, for these simulations we used $30\%$ of patients, learned qualitatively from Figure~\ref{fig:ESS_V_IA_plot_illustrative}. The maximum sample size was $80$. The optimal column is the result of enumerating over all possible IAs and choosing the one that minimises the sample size. }
    \centering
    \begin{tabular}{c | c | c | c}

        \hline

        \phantom & \multicolumn{3}{c}{Type of IA placement} \\

        \toprule
        
        Number of IAs & Optimal IA & Equally spaced IA & Equally spaced after waiting period IA \\

        \midrule
        
        One IA & 66.2 & 68.9 & 66.7 \\
        Two IA & 60.1 & 63.3 & 61.4 \\
        Three IA & 56.8 & 60.0 & 58.8 \\

        \bottomrule
        
    \end{tabular}

    \label{tab:my_label_table_1}
\end{table}

Table~\ref{tab:my_label_table_1} presents the ESS for the optimal IA placement for $1$, $2$, and $3$ IA with a maximum sample size of $80$ and the standard $\theta_{0} = 0.2$ and $\theta_{1} = 0.4$. This has been calculated  via Simon (1989)'s \cite{simon1989} enumeration method, calculating the ESS at every single IA and then returning the best. Note that due to computational constraints, $\lambda$ and $\gamma$ have not been optimised for every simulation, but type I error is still controlled below $10\%$ (exact values for $\lambda$ and $\gamma$ in the different simulations can be found in the code). Therefore, this table is purely illustrative, aiming to compare the ESS improvement from optimising the IA placement rather than simply equally spacing them.

We can see in Table 1 that there is significant gain when we have a waiting period at the start of the trial. Waiting period in this context is similar to a burn in that would be used for other adaptive methods, where there is a portion of the trial we do not consider when we are equally spacing. Here we wait for the first $30\%$ of patients, based on the flat plateau at the start of Figures such as Figure~\ref{fig:ESS_V_IA_plot_illustrative} (there is no reason to have an IA when it has minimal chance of stopping). Further, the ESS is again improved when we optimise the IA, matching the results of He et al. (2025) \cite{He2025}. This table therefore shows that even a preliminary analysis like Figure~\ref{fig:ESS_V_IA_plot_illustrative} is able to provide useful information about the period of time where an IA is unlikely to have an effect. In addition to this, Table~\ref{tab:my_label_table_1} also shows that there is indeed diminishing returns to be had from increasing the number of IA on the ESS. This work could also be replicated for other trial characteristics of interest, such as power and expected number of successes to illustrate the effect that IA can have on those. Further you could consider jointly optimising these characteristics as long as you are able to specify a utility function, but such work falls beyond the scope of this paper. Finally, this table also offers guidance on where to allocate resources during the trial planning. For larger and complex trials, it can be computationally unfeasible to optimise via enumeration, and therefore a table like this for a simpler example could provide key information on whether the ESS improvement from optimising versus equally spacing justifies the other practical considerations.

\section{Response-Adaptive Randomisation}
\label{sec:RAR}

Response-Adaptive Randomisation (RAR) is an alternative randomisation scheme to fixed randomisation. Fixed randomisation randomises patients based on a set proportion decided in the trial protocol: typically patients would be equally allocated amongst all the treatment arms, such as in the original BOP2 papers \cite{zhou2017bop2} \cite{Zhao2022}. On the other hand, RAR adapts the randomisation based on the data observed so far and uses that information to update the randomisation scheme, hoping to improve select operating characteristics. This section will be taking the RAR originally defined by Thompson (1933) \cite{thompson1933} and applying it to BOP2 to see the effect it has on the proportion of patients allocated to the superior treatment, as well as any effect it may have on power, to help justify to the practitioner whether it would be appropriate to consider for their trial. Unlike the previous section where we optimised the IA placement, IA in this section is equally spaced unless stated until Section~\ref{sec:RARandIA} (this is so we can isolate the effect RAR has by itself before considering IA and RAR jointly in Section~\ref{sec:RARandIA}).

\subsection{Two-arm}

Recall in Section~\ref{sec:BOP2} that the decision boundaries are based on posterior probabilities $P( {\theta}_{1} \geq {\theta}_{0} | D ) $. This makes Bayesian RAR, BRAR, efficient to implement in the BOP2 setting as it uses the very same probabilities. Therefore if we update the BRAR at every IA, we incur no additional computational cost from calculating the posteriors. 

The posterior probability reflects our confidence in one treatment being better than the other, however using this by itself for the allocation probabilities can lead to volatile and underpowered results \cite{thall2007}. That is why it is common to consider methods that slow down the speed of adaptation but still maintain the allocation to the superior arm: tuning, clipping, and burn-in, which we now define.

The burn-in refers to a period of time before the BRAR begins to adapt the randomisation. Before that, equal fixed allocation is used. This allows a minimum number of patients to be allocated to the control and the experimental treatment before BRAR can potentially skew the allocation to far in one direction. Having a burn-in prevents an analysis where most of the data is on one arm the patients are not split up enough to do a valid analysis \cite{ware1989}. For our analyses, we will be using a burn-in that goes up until the first IA, which for the analysis in this section is $20$ out of $80$ patients. Tang et al. (2025) \cite{tang2025} provides a more detailed analysis on burn-in placement and the effect it can have on a two-arm response-adaptive trial.

Meanwhile, Thall and Wathen (2007) \cite{thall2007} demonstrates the effect that tuning can have on the allocation. It involves the allocation probability being proportional to $P( {\theta}_{1} \geq {\theta}_{0} | D )^c $, where $c$ is some tuning parameter that pulls the allocation towards equal randomisation. We take $c = \frac{n}{2N}$, which has the intuitive behaviour that the allocation becomes more aggressive the more total information we have in the study.

Like burn-in and tuning, clipping prevents allocation probabilities from becoming too extreme. In this work, we do not use clipping as our tuning generally prevents the allocations from becoming too close to $0$ or $1$.

Formally then, our allocation probability begins with equal allocation, before being updated at every interim to

\[ p_{1} = \frac{P( \theta_{1} \geq \theta_{0} | D )^{c}}{P( \theta_{1} \geq \theta_{0} | D )^{c} + P( \theta_{1} < \theta_{0} | D )^{c}}, \]

where $c = \frac{n}{2N}$, and $p_{1}$ represents the probability of allocating the next patient to the experimental treatment (likewise $p_{0}$ the probability of allocating to the control).

We use this BRAR scheme and apply it to the BOP2 design of Zhao et al. (2022) \cite{Zhao2022}. Our simulations used a maximum sample size of $80$ patients, an IA carried out every $20$ patients, a control success rate of $0.2$, and parameters ( $\lambda = 0.91$, $\gamma = 0.93$ for equal randomisation, $\lambda = 0.9$, $\gamma = 0.86$ for BRAR) to control the type I error rate at $0.1$ and maximise power for an experimental treatment success rate of $0.4$. Table 2 shows the results of these $10,000$ simulations. 

\begin{table}[ht!]
    \centering
            \caption{Table showing the differences between fixed randomisation and BRAR for the two-arm BOP2. This was carried out with a control effect of $0.2$ and varying treatment effect $\theta_1$. The maximum sample size was $80$ and interims were carried out every $20$ patients. Power represents rate of rejection of the null hypothesis, ESS is Expected Sample Size, and Prop is the proportion of patients allocated to the treatment arm on average. }
    
            \begin{tabular}{c | c | c | c | c | c | c }
    
                \toprule   
                 &  \multicolumn{3}{|c|}{Equal randomisation} &  \multicolumn{3}{|c|}{Bayesian RAR} \\

                \toprule  
                
                \textbf{$\theta_1$} & \textbf{Power} & \textbf{ESS} & \textbf{Prop} & \textbf{Power} & \textbf{ESS} & \textbf{Prop} \\
    
                \midrule
                
                0.1 & 0.005 & 36.2 & 0.5 & 0.007 & 34.8 & 0.499 \\
                $H_{0} $: 0.2 & 0.086 & 51.0 & 0.5 & 0.097 & 49.4 & 0.523 \\
                0.3 & 0.372 & 60.2 & 0.5 & 0.381 & 58.6 & 0.560 \\           
                $H_{1} $: 0.4 & 0.728 & 59.6 & 0.5 & 0.713 & 59.0 & 0.588 \\
    
                \bottomrule
                
            \end{tabular}

    \label{tab:my_label2}

\end{table}

The clearest trend from Table \ref{tab:my_label2} is that BRAR improves on the number of patients allocated to experimental treatment whilst decreasing the power slightly when $\theta_1 = 0.4$. This is a common trade-off that can come as a result of RAR: balancing the benefit for the patients in the trial by increasing the chance they receive the best treatment, versus maximising the power of choosing the best treatment (and thus maximising the treatment quality for patients after the trial). For many rare diseases, the patient horizon may be small enough to justify a small power loss in exchange for more patients in the trial receiving the best treatment.

Table 2 also shows that the ESS is consistently lower across every scenario for BRAR than for equal randomisation, which means that fewer patients are getting being enrolled in the trial so quicker and more cost effective trials. On the other hand, in large patient populations it may be more reasonable to maximise the power to allow future patients a greater chance of receiving the best treatment. Table 2 tells us that there are advantages and disadvantages to BRAR in the BOP2 context, and it depends on the nuances of a specific trial to determine what trade-offs you are willing to make and what operating characteristics you wish to prioritise.

\begin{figure}[ht!]
            \centering

            \caption{Plot showcasing different average proportion of patients allocated to the treatment arm with a control success rate of $0.1$ (top plot) and $0.5$ (bottom plot) as the treatment success rate varies. The red line signifies $50\%$. }
            
            \includegraphics[width=1\linewidth]{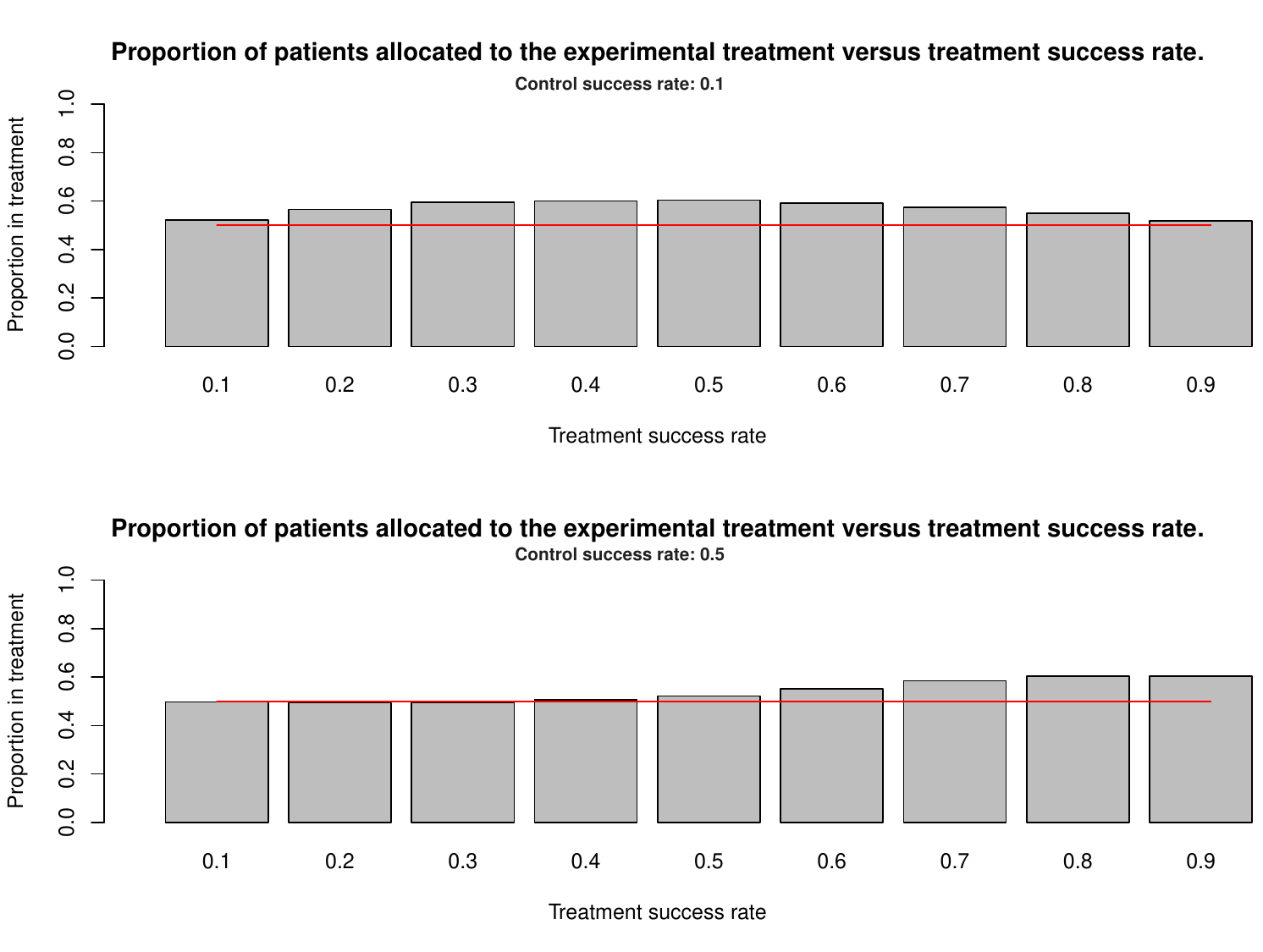}

            \label{fig:bar_chart_2_new}
\end{figure}

The goal of BRAR is to allocate more patients to the superior treatment. However, we can see in Table 2 that this design fails to do that for $\theta_{1} = 0.1$, allocating roughly equal proportions to both arms despite the control ($\theta_1 = 0.2$) having a higher success rate. We also see for $\theta_{1} = 0.2$ that more patients are allocated to the experimental treatment on average, despite the arms having identical success rates. Figure~\ref{fig:bar_chart_2_new} is consistent with these finding as well, showing that when the control outperforms the experimental treatment BRAR tends not to allocate more patients to the control arm.

\begin{figure}[ht]
    \centering
    \includegraphics[width=\linewidth]{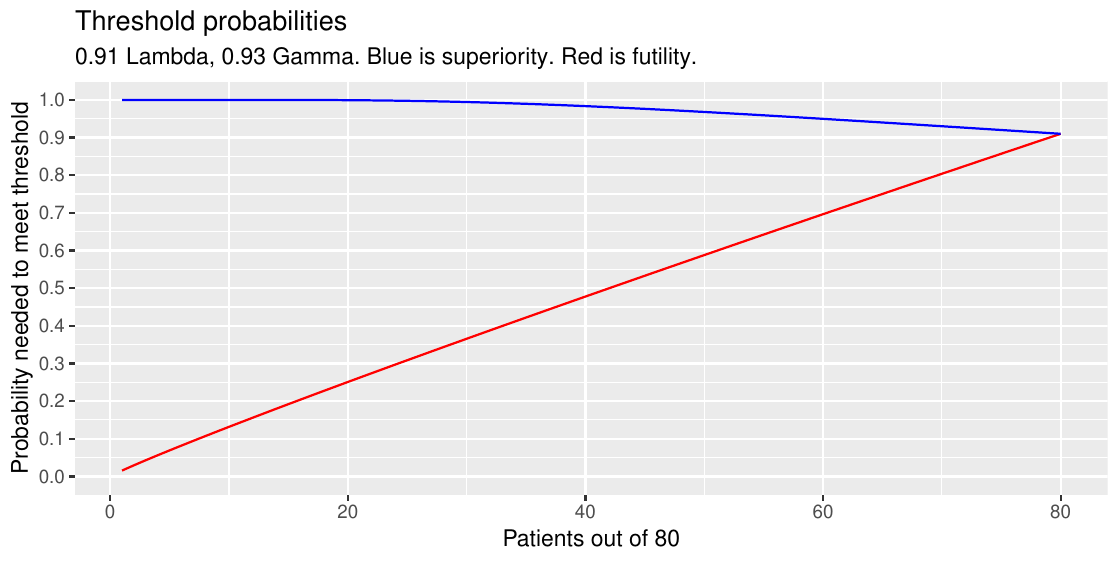}
    \caption{Plot showcasing the asymmetry between the BOP2 futility boundaries. Notice how futility rises significantly faster than superiority falls, showing how it is easier to stop for futility. }
    \label{fig:Threshold_plot}
\end{figure}

Figure~\ref{fig:bar_chart_2_new} shows that $p_{1}$ tends to $0.5$ when $\theta_{0} > \theta_{1}$: when the control arm performs better the BRAR tends to equal allocation. Furthermore, we see that when the control success rate is identical to the treatment success rate, $\theta_{0} = \theta_{1}$ , the design prefers the treatment and allocates more patients to the treatment on average, $p_{1} > 0.5$. Finally, when $\theta_{1} >> \theta_{0}$, the allocation probability $p_{1}$ tends to $0.5$: equal randomisation (for example, $\theta_1 = 0.9$ is closer to $0.5$ proportion than $\theta_1 = 0.8$ for $\theta_0 = 0.1$). These observations are all a result of how the asymmetric decision boundaries work with BRAR. Recall in Section~\ref{sec:BOP2} (and Figure~\ref{fig:Threshold_plot}) that it is easier to stop for futility than it is to stop for efficacy. This means that when the control outperforms the experimental treatment, we are likely to end early, but when the treatment outperforms the control, we are likely going to need to collect more data. As our burn-in coincides with the first IA, a control with a greater success rate than the experimental treatment will rarely not stop for futility at the first IA, meaning that the trial never progresses to using BRAR and thus the proportion plateaus at $0.5$. Similarly, in the case when the experimental treatment has the same success rate as the control, the trial ends early for efficacy when the control does well by chance, but when the treatment performs well the trial continues and RAR begins to allocate, prioritising the treatment because it has performed well. Finally, we see the plateau as we increase our treatment success even more as it finally becomes effective enough to end the trial early for efficacy at the first IA, preventing any adaptive randomisation once again. This illustrates how relatively simple and well studied RAR designs such as BRAR can lead to unforeseen complications and nuances when interacting with other aspects of the trial. Whilst not necessarily negative in all contexts, these results may be undesirable but can simply be prevented by having symmetric decision boundaries or stricter RAR tuning. We pick this point up again in the discussion: how to handle and evaluate unexpected effects of RAR.

\subsection{Multi-arm}

BRAR schemes in multi-arm designs become more varied and more complicated due to the additional arms they need to consider. This paper will be looking at two schemes which have been discussed in the literature. First, the Trippa scheme \cite{trippa2012} works by protecting allocation to the control arm, and then allocating based on which treatment is performing best out of the rest:

\[ p_{0} \propto  \frac{1}{n} \exp ( \max_{i} (n_{i}) - n_{0})^{c}, \quad p_{i} \propto P(\theta_{i} >  \Theta)^{c},\]

where $\Theta$ is a pre-specified threshold of meaningful efficacy, $p_{i}$ denotes the allocation probability to the $i$th arm, and we are still taking the tuning parameter  $c = \frac{n}{2N}$ . A key observation with this Trippa design is that it does not consider $\theta_{0}$ at any point (although modifications do exist which replace $\Theta$ with $\theta_{0}$ \cite{das2024implementing}). This means that the performance of the control arm does not effect how many patients are allocated to the control arm. It does this to preserve the least power: ensuring that the control arm has a comparable number of patients to the most prevalent treatment arm. Thus, the Trippa scheme aims to optimise least power and patients allocated to the best experimental treatment at the cost of being unable to allocate fewer patients to an unsuccessful control (or more patients to a successful control). Additionally, the computational cost of the Trippa scheme is very low: it only requires calculations of a Beta cumulative distribution function. 

This computational cost is in contrast to the other scheme we will considering, the Maximum BRAR scheme. Discussed  in Kaddaj et al. (2024) \cite{kaddaj2024}, this scheme is a more natural extension to the two-arm BRAR scheme: allocating to an arm based on the probability that the treatment is the best. Formally, we have

\[ P(\text{treatment i}) \propto P( \theta_{i} \geq \theta_{j}, \forall j \neq i | D )^{c}. \]

Unlike the Trippa scheme, calculating this posterior is computationally taxing. The exact formula of Miller (2015) \cite{miller2015} does provide an exact formula for $4$ arms or fewer, although it is recursive. Beyond $4$ arms, empirical approximation of the posterior c.d.f is required, which is both slow and inaccurate \cite{kaddaj2024}. Despite this, the Maximum scheme has more flexibility in handling extreme controls ($\theta_{0} \approx 0$ or $\theta_{0} \approx 1$), naturally deals with the absence of a control, and is more likely to be able to allocate a larger proportion of patients to the superior treatment arm due to not protecting allocation to the control arm. 

Figure~\ref{fig:alloprobmax_new} in the appendix illustrates the differences between the two RAR schemes. We can see that the Trippa scheme can not exceed the probability of allocation to the control, despite the treatment outperforming it with $30$ successes versus $13$ (out of $30$). On the other hand, the Maximum scheme shows the intuitive behaviour we would want from a RAR scheme: the allocation is greatest on the arm which is performing the best.

To compare the schemes quantitively, we will be considering $4$ different scenarios. Scenario $1$ will be be the controlled Least Favourable Configuration (LFC) with $4$ arms. This means that there will be a singular effective arm ($\theta_1 = 0.6$, $\phi_{1} = 0.2$) against $3$ equally ineffective arms, including the control ($\theta_{0,2,3} = 0.45$, toxicity rate = $0.3$). Scenario $2$ will be a $4$ arm controlled staircase: each treatment is slightly more effective than the previous one (so $\theta_{0} = 0.45,\theta_{2} = 0.55, \theta_{3} = 0.6, \theta_{1} = 0.65$, and the toxicity rate is $0.3,0.25,0.2,0.15$ respectively). Scenario $3$ will be the uncontrolled LFC and Scenario $4$ will be the uncontrolled staircase, both $3$ armed designs equivalent to their controlled counterparts but without a control. Table \ref{tab:scenarios} summarises these scenarios:

\begin{table}
    
    \centering

    \caption{Table highlighting the different scenarios being considered. Recall, $\theta_{i}$ and $\phi_{i}$ refer to the success and toxicity rate of treatment $i$ respectively. }
    
    \begin{tabular}{|c|c|}\hline
        Scenario 1 & $(\theta_{0}, \theta_{1}, \theta_{2}, \theta_{3}) = (0.45,0.6,0.45,0.45) $, 
        $(\phi_{0}, \phi_{1}, \phi_{2}, \phi_{3}) = (0.3,0.2,0.3,0.3) $\\\hline
        Scenario 2 & $(\theta_{0}, \theta_{1}, \theta_{2}, \theta_{3}) = (0.45,0.65,0.55,0.6) $, 
        $(\phi_{0}, \phi_{1}, \phi_{2}, \phi_{3}) = (0.3,0.15,0.25,0.2)$\\\hline
        Scenario 3 & $(\theta_{1}, \theta_{2}, \theta_{3}) = (0.6,0.45,0.45) $, 
        $(\phi_{1}, \phi_{2}, \phi_{3}) = (0.2,0.3,0.3)$\\\hline
        Scenario 4 & $( \theta_{1}, \theta_{2}, \theta_{3}) = (0.65,0.55,0.6) $, 
        $ (\phi_{1}, \phi_{2}, \phi_{3}) = (0.15,0.25,0.2)$ \\ \hline
    \end{tabular}
    
    \label{tab:scenarios}
\end{table}

All scenarios have a maximum sample size of $60$ per arm (so $240$ for scenarios $1$ and $2$ and $180$ for scenarios $3$ and $4$) and have an IA every $60$ patients. Recall that $\lambda$ and $\gamma$ have been optimised for every scenario with every scheme to protect the FWER at $10\%$ and maximise the least power. Exact values for $\lambda$ and $\gamma$ can be found in the code freely available with this paper.

Table \ref{tab:my_label3_new} contains comparisons of the different RAR schemes for Scenario 1 and 2. We compare the Trippa design and the Maximum design against the original multi-arm BOP2 design of Mulier et al. (2024) \cite{Mulier2024}, and a Reallocation design. The Reallocation design is identical to the design of Mulier et al. (2024) \cite{Mulier2024}, except when an arm is ended early for futility the patients that would have been allocated to that arm are equally reallocated to the other arms, as opposed to reducing the sample size of the trial. Further, in Table 3 we consider the treatment ESS rather than the ESS of the trial. This is the ESS of the effective treatment arm, so that we can effectively compare how well the schemes perform in sending patients to the most effective arms. 

\begin{table}
    \centering

    \caption{Table showing the operating characteristics from scenarios $1$ and $2$. It compares least power, ESS in the effective treatment arm, the probability of the effective arm stopping early for futility, and the proportion of patients in the most effective arm against $4$ different BOP2 designs. }
    
    \begin{tabular}{c | cccc}
    \toprule
         1: Controlled LFC & \textbf{Least Power} & \textbf{Treatment ESS} & \textbf{Early Stopping} & \textbf{Prop} \\

    \midrule
         
        Mulier design & 0.46 & 48.1 & 0.40 & $30\%$\\
        Reallocation design & 0.55 & 66.9 & 0.37 & $34\%$\\
        Trippa design & 0.53 & 63.7 & 0.31 & $31\%$\\
        Maximum design & 0.55 & 86.2 & 0.31 & $40\%$\\

    \toprule

        2: Controlled Staircase & \textbf{Least Power} & \textbf{Treatment ESS} & \textbf{Early Stopping} & \textbf{Prop} \\

    \midrule
         
        Mulier design & 0.74 & 54.4 & 0.18 & $27\%$\\
        Reallocation design & 0.78 & 64.8 & 0.18 & $28\%$\\
        Trippa design & 0.78 & 56.5 & 0.15 & $24\%$\\
        Maximum design & 0.78 & 80.7 & 0.15 & $34\%$\\
         
    \bottomrule 

    \toprule
         3: Uncontrolled LFC & \textbf{Least Power} & \textbf{Treatment ESS} & \textbf{Early Stopping} & \textbf{Prop} \\

    \midrule
         
        Mulier design & 0.75 & 56.0 & 0.15 & $44\%$\\
        Reallocation design & 0.84 & 93.8 & 0.12 & $54\%$\\
        Trippa design & 0.85 & 94.8 & 0.11 & $54\%$\\
        Maximum design & 0.85 & 100.0 & 0.10 & $57\%$\\

    \toprule
         4: Uncontrolled Staircase & \textbf{Least Power} & \textbf{Treatment ESS} & \textbf{Early Stopping} & \textbf{Prop} \\

    \midrule
         
        Mulier design & 0.96 & 59.2 & 0.028 & $36\%$\\
        Reallocation design & 0.96 & 67.5 & 0.026 & $38\%$\\
        Trippa design & 0.96 & 67.9 & 0.025 & $38\%$\\
        Maximum design & 0.96 & 75.9 & 0.024 & $42\%$\\
         
    \bottomrule

    \end{tabular}
    
    \label{tab:my_label3_new}
\end{table}

We can see in Table \ref{tab:my_label3_new} that the Maximum design consistently allocates the most patients, both in proportion and ESS, to the most effective treatment in both scenarios $1$ and $2$. On the other hand, the Trippa scheme allocates fewer patients to the most effective arm than the equal randomisation Reallocation design, simply because it tends not to allocate patients to non-control arms as seen in Figure~\ref{fig:alloprobmax_new}. The Trippa scheme also has a lower least power in Scenario 1, but mostly the least power is consistent between the non Mulier schemes (which is to be expected as the Mulier scheme will always see the fewest number of patients). In general, Table \ref{tab:my_label3_new} shows that the Maximum design either has the best or joint best operating characteristics for scenarios 1 and 2 across all the operating characteristics we consider. 

Furthermore, Table \ref{tab:my_label3_new} provides additional evidence that the Maximum design is the best choice out of the designs we consider. Even in the uncontrolled scenarios $3$ and $4$, the Maximum design has the highest power, treatment ESS, and proportion of patients allocated to the most effective treatment. It also has the smallest chance of erroneously stopping the most effective arm early for futility. Notably, this is a contrast to Table 2, where we observed in the two-arm setting that there was a trade off between power and proportion when using RAR. In the multi-arm setting, RAR can help randomise patients towards the most promising arms, maximising information where it would be most valuable to learn more, whilst randomising less towards the unpromising arms. Unlike the two-arm setting, this does not necessarily come at the cost of significantly fewer patients being allocated towards the control. Instead more patients come from other less effective arms, hence why it has a smaller impact on the power, or can even improve it as we see in Scenario 3.

\subsection{RAR and IA}
\label{sec:RARandIA}

\begin{figure}[ht!]
            \centering
            
            \caption{Plot showcasing how the proportion of patients allocated to a superior treatment arm in a RAR BOP2 design changes as the IA placement changes. The shaded area is an approximate $95\%$ confidence interval, calculated by taking the $2.5\%$ and $97.5\%$ quantiles of the $10,000$ simulations. }
            
            \includegraphics[width=1\linewidth, height = 0.8\linewidth]{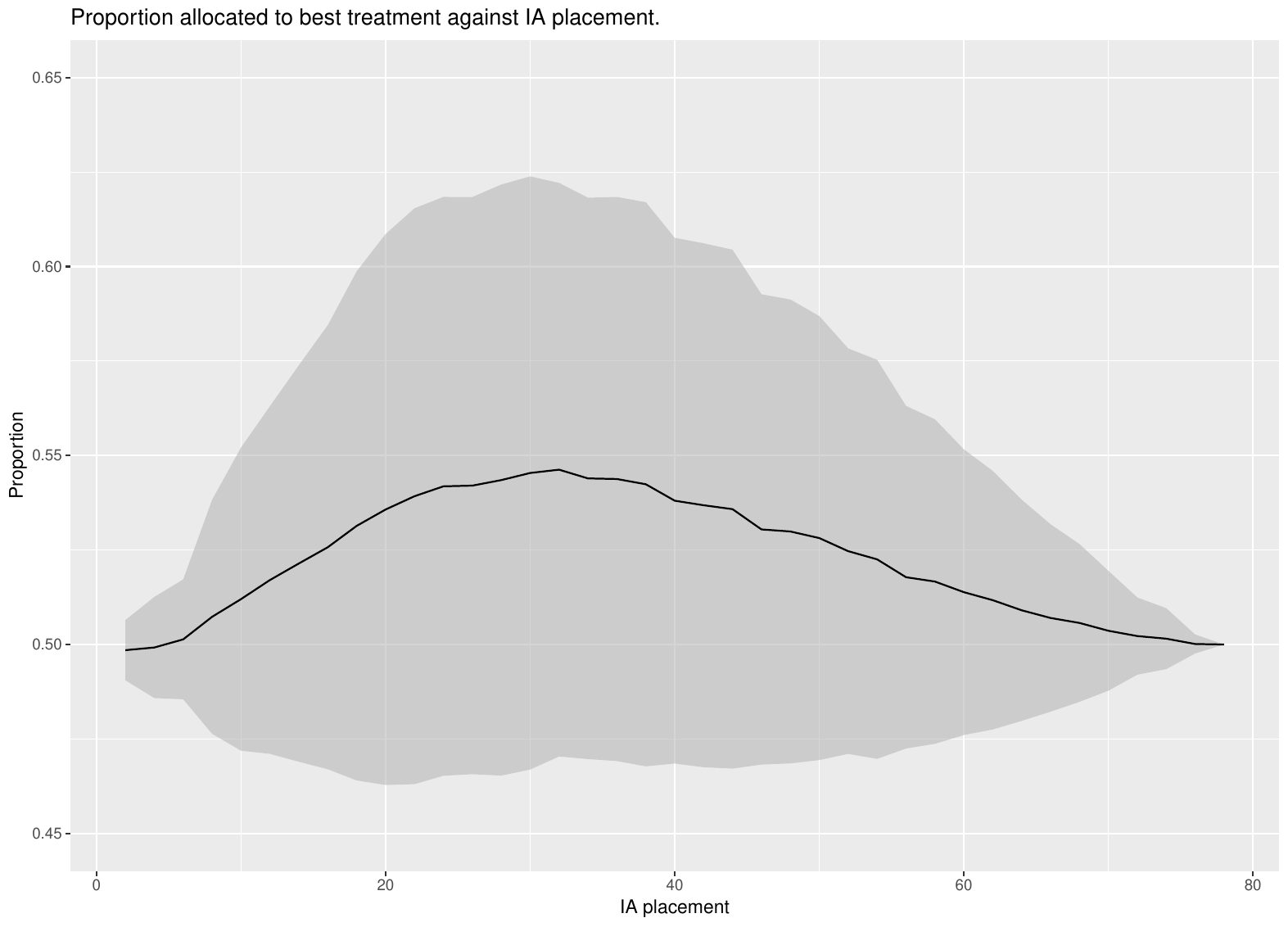}
            
            \label{fig:abc0}
\end{figure}

The RAR procedure we use does not update until the first IA. Therefore, it is natural to assume that the IA placement will affect the potential value of the RAR. This can be observed in Figure~\ref{fig:abc0}, where both the proportion of patients allocated to the superior treatment and the variance of the allocation proportion increases as we move our IA towards the region between $30$ to $40$. This again hows the trade off with RAR: optimising patients allocated to the superior treatment increases the variance of the final proportions, with a chance that a majority of patients will be allocated to the inferior arm. However, all of the $95\%$ confidence intervals are above only a $0.45$ proportion of patients allocated to the superior treatment, giving us confidence that if the RAR does misallocate it should not be a large misallocation.

Figure~\ref{fig:abc0} motivates the question of what is the optimal IA for maximising the proportion of patients allocated to the superior treatment arm. This builds on Table \ref{tab:my_label_table_1} in Section~\ref{sec:IA}, where we optimised the IA to see the effect it has on the ESS. Table 5 instead optimises the IA placement for the proportion of patients allocated to the superior treatment. Unlike Table \ref{tab:my_label_table_1}, not placing any IA in the first $30\%$ of patients (waiting period) and then equally spacing, which provided noticeable benefits in Table \ref{tab:my_label_table_1} is now lowering the proportion versus equal spacing. This corroborates Figure~\ref{fig:abc0}, where unlike the ESS plots where there is a plateau at the start, the proportion immediately starts increasing. This means that optimising for the proportion of patients on the superior treatment is a fundamentally different problem from minimising the ESS and will once again come down to practitioner trade-off on what is best to optimise for a specific trial design.

\begin{table}[ht!]
    \centering

    \caption{Table showing how the proportion to patients allocated to the superior treatment can change when we place our IA using different methods, using RAR. The waiting period used for this was $30\%$ of patients, learned qualitatively from Figure~\ref{fig:alloprobmax_new}. }
    
    \begin{tabular}{c | c | c | c}

        \hline

        \phantom & \multicolumn{3}{c}{Type of IA placement} \\

        \toprule
        
        Number of IAs & Optimal & Equally spaced & Equally spaced after waiting period. \\

        \midrule
        
        One IA & 0.546 & 0.539 & 0.525 \\
        Two IA & 0.554 & 0.548 & 0.532 \\
        Three IA & 0.554 & 0.547 & 0.532 \\

        \bottomrule
        
    \end{tabular}
    
    \label{tab:my_labeltable}
\end{table}

Additionally, we see the diminishing returns effect we saw in Table \ref{tab:my_label_table_1} much quicker here. Whilst the jump from one to two IA improves the proportion allocated to the superior treatment by $0.01$, there is no observable improvement from two to three IA (to three significant figures). Hence updating the randomisation more often does not necessarily lead to better trial characteristics, and supports the argument for updating the RAR at IAs rather than a fully sequential RAR design if additional updates do not yield significant improvements.

\section{Discussion}
\label{sec:Dis}

The BOP2 design introduced by Zhou et al. (2017) \cite{zhou2017bop2} is a powerful and flexible Bayesian framework. It can be extended to many different scenarios: multi-arm trials \cite{Zhou2020BOP2}, delayed endpoints \cite{zhao2023bop2}, and complex decision making \cite{Mulier2024}. The goal of this paper was to illustrate how adapting BOP2 to different design can significantly change operating characteristics, and thus how important it is to investigate and optimise trial decisions. Specifically, this paper investigates optimising the interim analysis placement and the randomisation schemes. Our results show that optimising IA placement can lead to improved ESS with minimal impact on other operating characteristics, whilst implementing RAR always improved the mean proportion of patients allocated to the superior treatment at the slight cost of power in the two-arm setting. On the other hand, it both improved both least power and proportion of patients allocated to the superior treatment in the multi arm setting, showcasing that BRAR should be a strong consideration for many multi-arm trials. 

The improved ESS obtained by optimising IA placements can offer a strategically aligned framework for generating optimum high-quality randomised evidence in oncology trials. The proposed approach enables efficient early stopping for futility or overwhelming efficacy, thereby reducing patient exposure to ineffective treatments and accelerating the transition to confirmatory studies, key aims of Project FrontRunner. The incorporation of Bayesian response-adaptive randomisation further strengthens this alignment by allocating more patients to treatments demonstrating emerging benefit while preserving the integrity of randomised comparisons. The features of the proposed approach therefore can support FrontRunner’s mission to modernize methodology in earlier-line metastatic settings and promote trial designs that are both ethically responsive and scientifically rigorous.

There are additional challenges that come with optimising trial design however. For example, introducing additional IAs leads to more meetings of the Data Monitoring Committee (DMC), it means the data is unblinded more often, and it can slow down the trial if recruitment has to be paused waiting for the results of the IA. These can all incur additional costs or introduce unworkable biases into the data. Table 6 provides a summary of the advantages that can come with optimising IA placement versus heuristically placing IA. The key takeaway from this paper is that whilst changing trial parameters can improve operating characteristics, the criteria and motivation for optimisation will be different for every trial. It is therefore important to consider what stakeholder priorities are in order to design a successful trial.

\begin{table}[ht!]
    \centering

    \caption{A table showcasing the advantages of taking time to optimise your IA placement and frequency versus following heuristics to simply place them. }
    
    \begin{tabular}{p{8cm} | p{8cm}}
         \textbf{Optimal IAs} & \textbf{Heuristic IAs} \\

        \toprule
         
         -Leads to better ESS and smaller trials & -Takes much less time to implement in the planning stage\\

         -Demonstrates how robust the trial is to changes in IA & -Often will give comparable ESS following the heuristic outlined in the IA section \\

         -The heuristics may not work with different trial designs and adaptations & -Using the same IA as other trials make the designs and results innately more comparable.  \\

        \bottomrule
         
    \end{tabular}
    
    \label{tab:my_label8}
\end{table}

Aside from operational issues, there are additional statistical advantages and disadvantages to consider when using RAR. Robertson et al. (2023) \cite{robertson2023} describes issues that can arise when using RAR and how to mitigate for them, such as time trends, statistical inference, and the chance of allocating more patients to an inferior treatment. RAR has incredible flexibility, with a wide range of schemes that solve different optimisation problems and fulfil different criteria, but innovative designs introduce challenging problems. This paper provides the code and outlines the workflow that can be replicated in the trial planning process to decide whether RAR would improve the quality of a trial. We encourage practitioners to consider the randomisation scheme they use in trials critically to ensure patients receive the best possible care.

There are many ways the work in this paper could be meaningfully extended. Foremost, the methodology of taking a principal component of trial design, such as the randomisation, and evaluating how it effects the trial when we change it can be applied to any other aspect of a trial. For example, consider recruitment speed and the effects it can have. If it is slower than anticipated, trials run over schedule and incur extra costs \cite{mcdonald2006}. Missing data at interim analyses may also impact the quality of the decisions made  \cite{das2024implementing}. On the other hand, if recruiting is quicker than expected, by the time an IA has been carried out a majority of the patients have already been recruited and randomised, meaning that the IA will have negligible effect. To a similar effect, delayed endpoints also make effective IA placement more difficult, as a long delay to endpoint means that many patients who have already been randomised will not have their data available at the IA, possible impacting the quality of its decisions. Therefore looking at how recruitment speed and delayed endpoints affect trial properties, as well as the effect of influencing it in terms of number of trial centres, could be useful information for trial planning. In addition to recruitment speed, heterogenous patient populations (for example, where one subgroup responds well to treatment and one subgroup does not) and the effect of considering joint versus hierarchical versus single endpoints could all be investigated to benefit future trial planning.  

Secondly, there is potential benefit to be gained from considering RAR for safety driven non-inferiority trials. This paper has shown that the main benefit of RAR is maximising patient benefit by increasing allocation to the better treatment. However, if two treatments have comparable efficacies, then we would want to start allocating based on the better safety profile. There is no obvious way to use RAR such that it uses prioritises efficacy data but still uses safety data to determine allocation probabilities. Chiaruttini et al. (2025) and
Kanrar et al. (2025) \cite{chiaruttini2025} \cite{kanrar2025} both introduce novel RAR methods of using efficacy and safety data in conjunction to develop allocations for setting which prioritise efficacy, but further work is needed to optimise and compare these methods. This sets the foundation for non-inferiority trials in later phase dose finding, using RAR to ensure that patients receive a minimum level of established efficacy. There is space to research how allocating more resources to later phase dose finding as opposed to early phase dose finding, as is motivated by the FDA's Project Optimus \cite{optimus}, changes the drug development timeline. 

A final issue to consider are computational challenges. The algorithms derived by Miller (2015) \cite{miller2015} for efficient beta c.d.f calculation are quick, yet the calculation of posterior probabilities are still the bottleneck of the simulations, especially for more complex schemes like the Maximum BRAR from Kaddaj et al. (2024) \cite{kaddaj2024}. Struggling to compute the operating characteristics of complex trials can mean that novel ideas are not tested and patients may not be receiving the best possible form of care. Therefore, methods to ease computational complexity such as exact operating characteristic calculation as discussed in Baas et al. (2024 \cite{baas2024} are vital to ensuring continued progress in quality of clinical trials.

\section*{Code availability}

Reproducible code can be freely sourced at Github via https://github.com/Con1con2/BOP2-RAR-paper , any questions please direct them to Connor Fitchett.

\section*{Appendix}

The appendix contains supplementary figures, tables, and notes. This includes a table summarising all the notation present and figures illustrating the decision thresholds. 

A nuance on the confidence intervals (CI) in Figure~\ref{fig:ESS_V_IA_plot_illustrative}, which importantly tell us that spiky variation in power is not (entirely) due to Monte Carlo simulation error. The power plot CI is a simple approximate normal $95\%$ CI, taking the variance to be upper bounded at $0.5$ (the largest possible variance for a beta distributed random variable). On the other hand, because the normal approximation is less justified for the ESS as it is not symmetric in its distribution, especially when it gets close to $80$. The ESS can always go below the maximum sample size, but never above. Therefore, the top CI has been calculated via bootstrap: taking batches of the $100,000$ simulations to get ESS approximation, and then taking the $2.5\%$ and $97.5\%$ quantiles to get a $95\%$ bootstrap CI. This method would also generate a valid CI for the power, however it is less informative than the approximate normal CI.

\begin{table}[ht]
    \centering
    \begin{tabular}{p{2cm} | p{8cm}}
        \textbf{Notation} & \textbf{Description} \\

        \toprule
        
        $\theta_{i}$ & The Bayesian random variable representing the probability of the $i$th treatment arm being successful. $\theta_{0}$ denotes the control success probability. \\
        $\phi_{i}$ & The Bayesian random variable representing the probability of the $i$th treatment arm yielding a toxic response (independent of success). $\phi_{0}$ denotes the control toxicity probability. \\
        $p_{i}$ & The allocation probability representing the probability the next patient is allocated to the $i$th treatment arm. $p_{0}$ denotes the control allocation probability. \\
        $\pi$ & The Bayesian prior for $\theta$. \\
        $ D$ & Shorthand for the data collected so far in the study, used to denote posterior probabilities by conditioning on it. \\
        $ \lambda , \gamma$ & The parameters that determine the stopping boundaries of BOP2, optimised to protect type I error. \\
        $ \Phi() $ & The cumulative distribution function for the standard normal. \\
        $Z_{ \alpha}$ & The $\alpha$ quantile of the standard normal, that is to say $\Phi(Z_{\alpha} ) = \alpha$\\
        $n, N$ & $n$ represents the current sample sample size, $N$ represents the maximum sample size. \\
        $n_i$ & The current sample size of the $i$th arm. $n_{0}$ denotes the control sample size. \\
        $ \Theta $  & Denotes a pre-specified treatment effect that is constant (used in uncontrolled studies in place of the control) \\
         $c, \eta$ &  Tuning parameters used to determine the strength of RAR. \\
        $ K$ & $K$ is the maximum number of treatment arms in a multi-arm context. \\
        $a$ & $a$ is how many are still open to recruitment (have not been dropped for futility yet.) \\

        \bottomrule
         
    \end{tabular}
    \caption{Notation summary for expressions that appear throughout the paper. }
    \label{tab:my_label_notation. }
\end{table}

\newpage
\phantom{a}

\newpage
\phantom{a}

\begin{figure}[ht]
    \centering
    \includegraphics[width=1\linewidth, height = 0.8\linewidth]{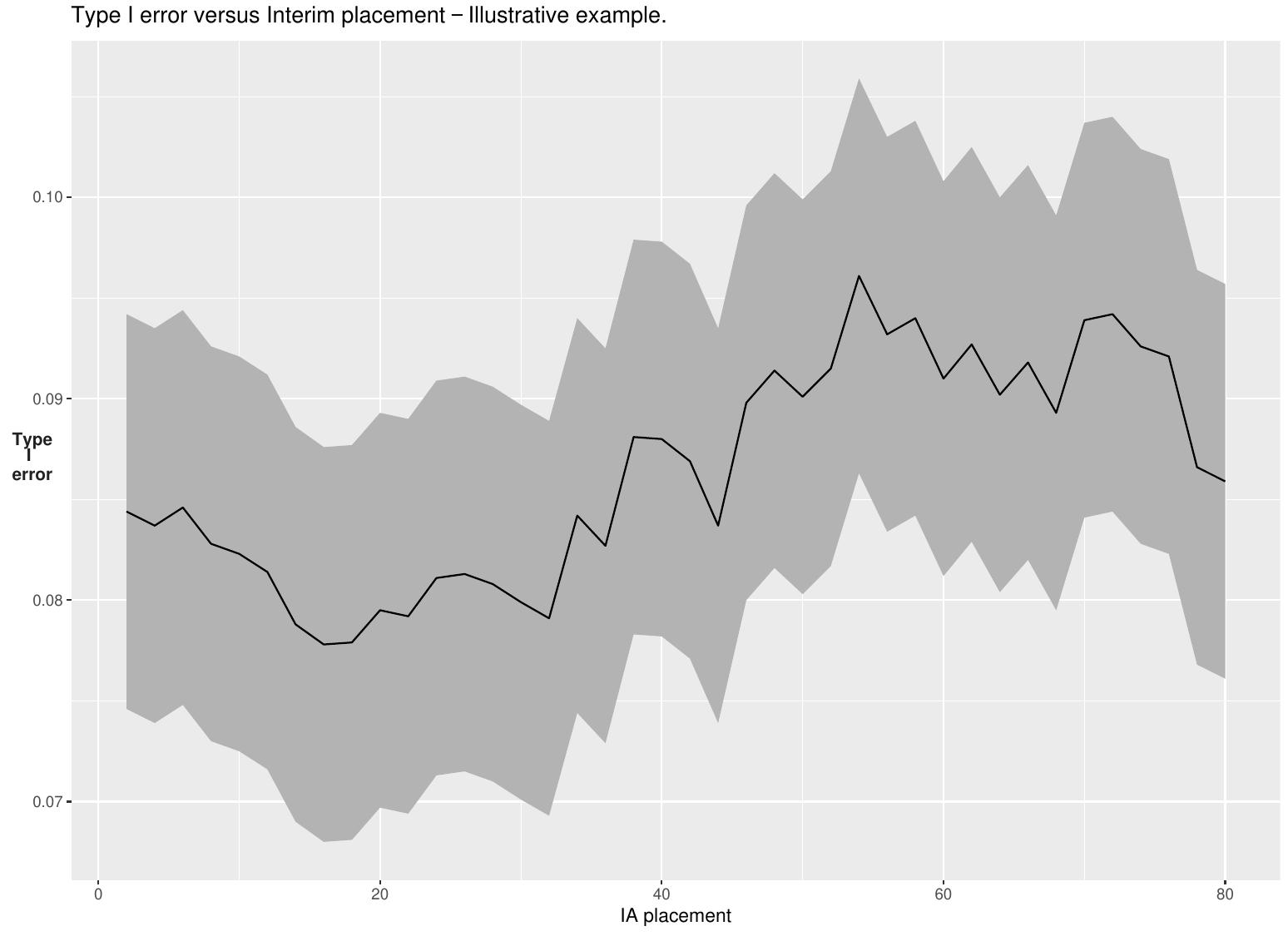}
    \caption{The accompanying type I error rate plot for Figure~\ref{fig:ESS_V_IA_plot_illustrative}, which was optimised at a IA at 60 patients and a final analysis at 80. Note how the type I error rate is always below the target $0.1$. }
    \label{fig:appen_2}
\end{figure}

\newpage
\phantom{a}

\begin{figure}[ht]
    \centering
    \includegraphics[width=1\linewidth, height = 0.8\linewidth]{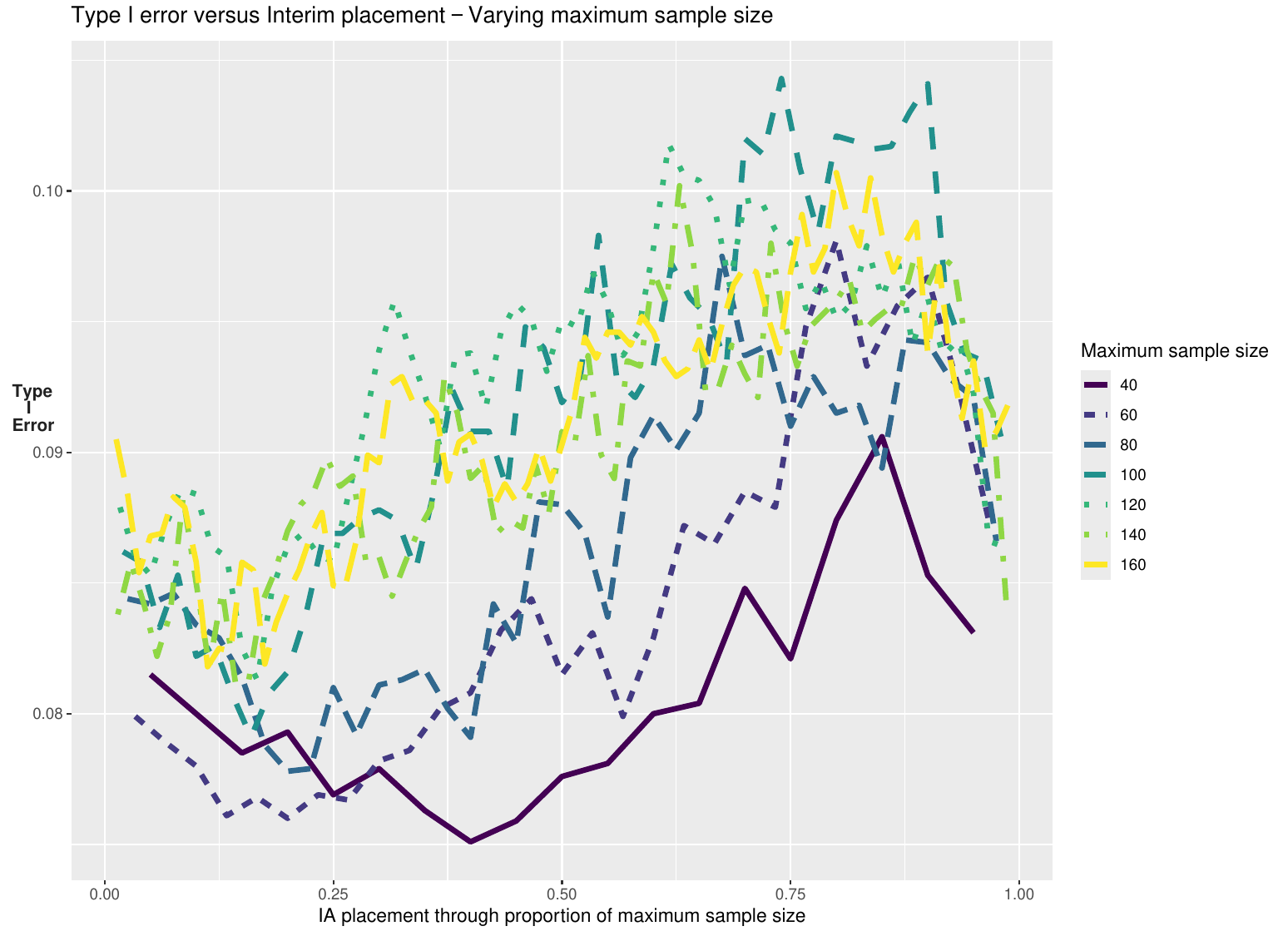}
    \caption{The accompanying type I error rate plot for Figure~\ref{fig:ESS_V_IA_plot_vary_ss}, which was optimised for 80 patients. The type I error rate is below $0.1$ for most of the maximum sample sizes, with a small inflation at for $100$ }
    \label{fig:appen_3}
\end{figure}

\newpage
\phantom{a}

\begin{figure}[ht]
    \centering
    \includegraphics[width=1\linewidth, height = 0.8\linewidth]{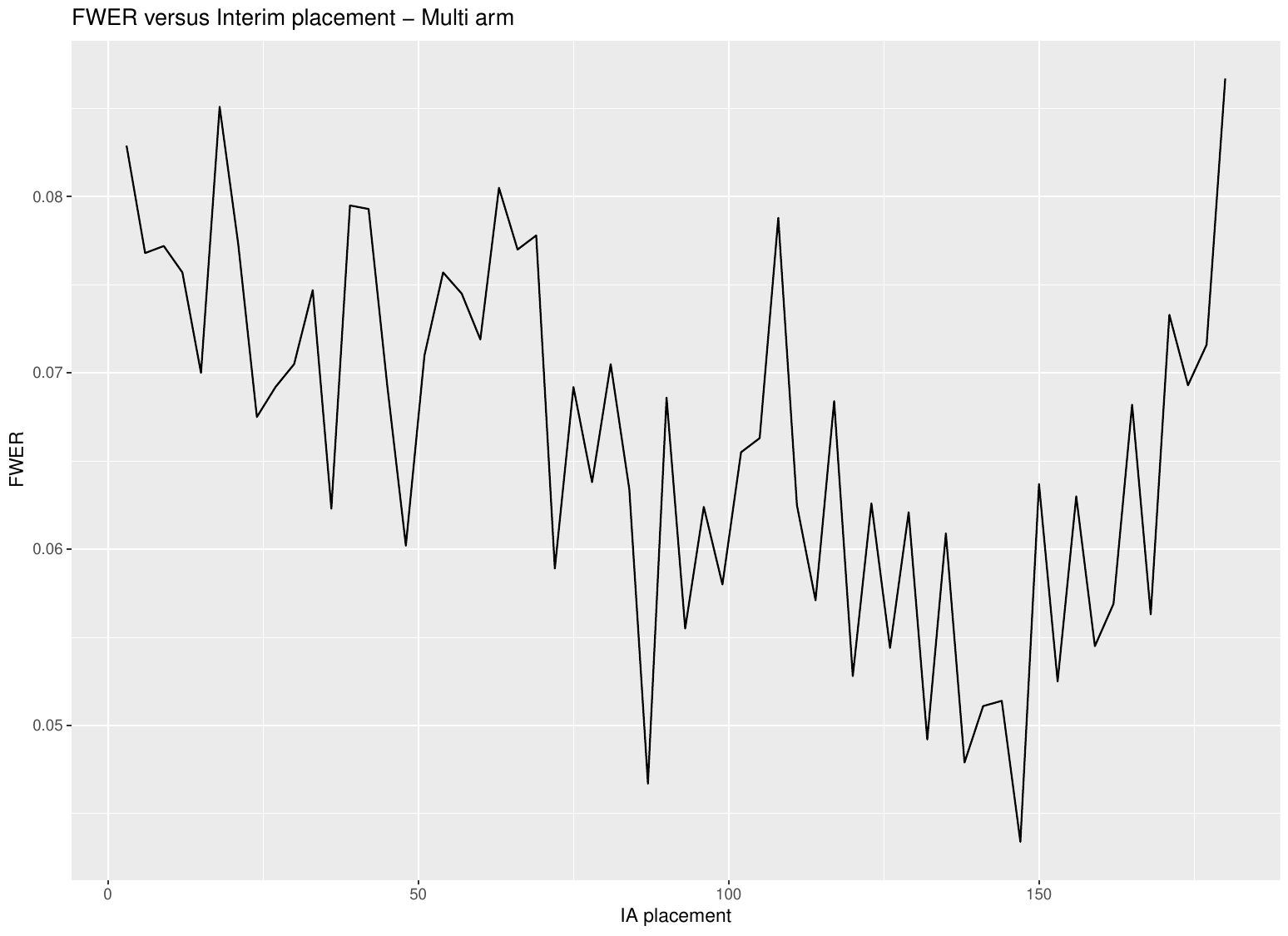}
    \caption{The accompanying FWER plot for Figure~\ref{fig:ESS_V_IA_plot_multi}, which was optimised for 180 total patients. The FWER is below $0.1$ for the entire plot. }
    \label{fig:appen_4}
\end{figure}

\newpage
\phantom{a}

\begin{figure}[ht!]
            \centering

            \caption{Plot showcasing how the allocation probability to the control arm (red) and the experimental treatment arm (blue) change as the treatment arm gets more successes, under the different RAR schemes. The green line is the number of successes observed in the treatment. Note that this is in a three arm setting, with the other treatment performing the same as the control arm, taken after $30$ patients per arm in an $180$ total patient trial. }
            
            \includegraphics[width=1\linewidth, height = 1\linewidth]{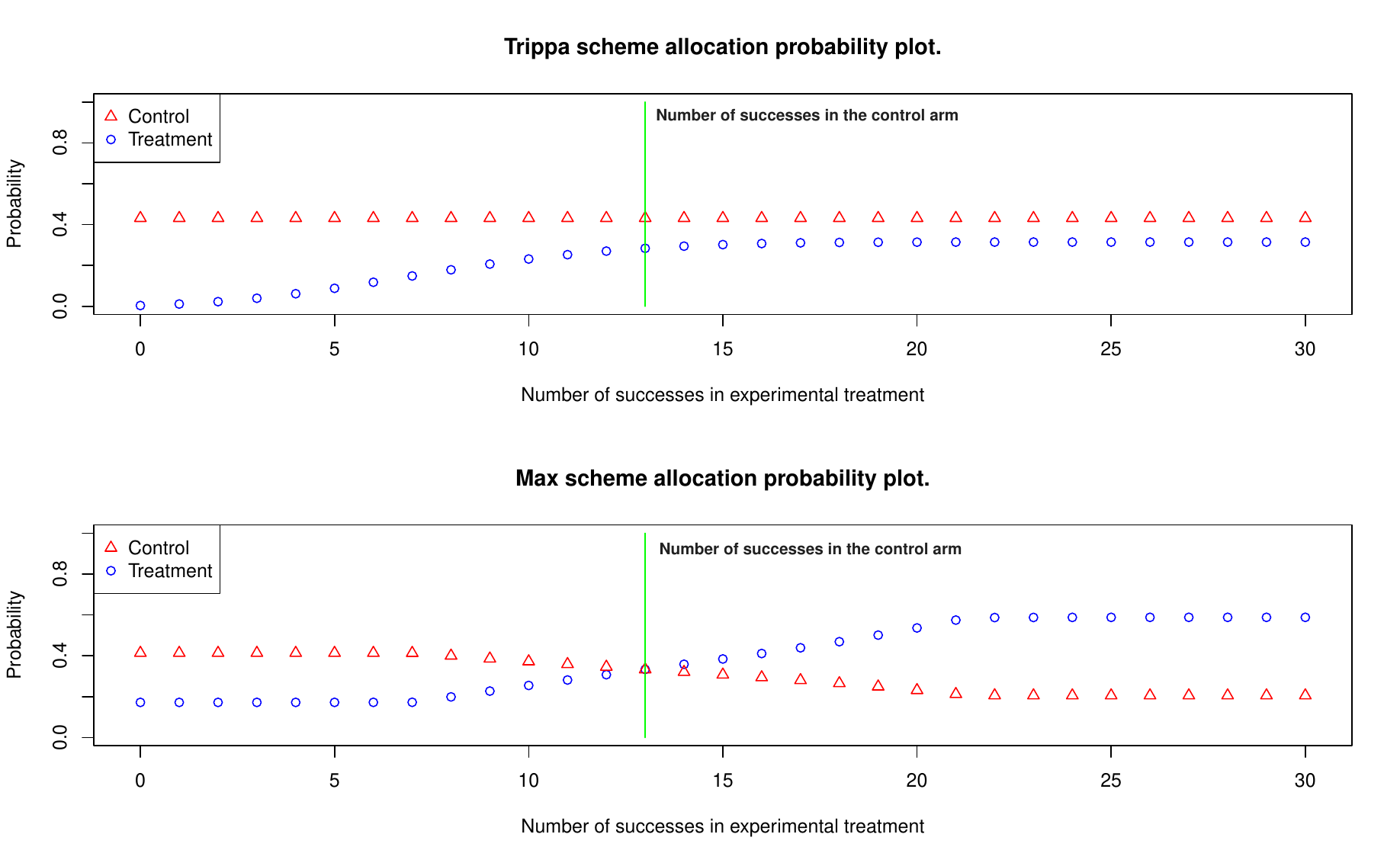}
            
            \label{fig:alloprobmax_new}
\end{figure}

\newpage
\phantom{a}
 
\newpage

\bibliographystyle{ieeetr}
\bibliography{refs} 

\end{document}